\documentclass{article}
\usepackage{amsmath}
\usepackage{amsfonts}
\usepackage{amssymb}
\usepackage[usenames]{color}
\usepackage{amsthm}
\usepackage{graphicx}
\usepackage[utf8]{inputenc}

\def\p{\partial}

\theoremstyle{remark}
\newcommand{\wt}{\widetilde}
\newcommand{\be}{\begin{equation}}
\newcommand{\ee}{\end{equation}}
\newcommand{\bea}{\begin{eqnarray}}
\newcommand{\eea}{\end{eqnarray}}
\newcommand{\beaa}{\begin{eqnarray*}}
\newcommand{\eeaa}{\end{eqnarray*}}

\newcommand{\nn}{\nonumber}
\renewcommand{\d}{\mathrm{d}}

\usepackage{authblk}
\begin{document}
\title
{Differential and other reductions of the self-dual conformal structure equations\thanks
{In memory of V.E. Zakharov}
}
\author{L.V. Bogdanov\thanks{leonid@itp.ac.ru}}
\affil{Landau Institute for Theoretical Physics RAS}
\date{January 22, 2025}
\maketitle
\begin{abstract}
The dispersionless integrable system we consider here 
was introduced to the literature rather recently,
it is connected with the general
local form of self-dual conformal structure (SDCS) for the signature (2,2).
In integrability  framework  this system possesses a rich structure 
of reductions, including differential reductions. We will discuss several characteristic reductions
for this system, using the Lax pair, hierarchy structure and the dressing scheme. 
We use reductions to construct solutions for the SDCS equations.
One of our goals is to present type B SDCS system and
consider its relations with the SDCS system.
\end{abstract}
\section{Introduction}
Differential reductions play an important role in the theory of integrable
systems. They represent some nonlinear differential relations compatible with 
the initial integrable system. A remarkable example of a differential reduction, connected
with the description of n-orthogonal surfaces,  
was  introduced to integrability by V.E. Zakharov \cite{Z98}.
N-orthogonal surfaces are characterised by the Darboux-Lam\'e system of equations. One part
of this system consists of the system of equations of conjugate surfaces
(diagonal curvature spaces), which is related
to N-wave equations, integrated  in 1974 by Zakharov and Shabat
\cite{ZS74}. Another part represents a set of 
compatible nonlinear differential relations connected with orthogonality, 
the meaning of which in the framework of the theory of
integrable systems was unclear.  V.E. Zakharov managed to find a description of
this reduction in terms of integrability scheme (the dressing scheme),
and it appeared rather simple, natural and elegant \cite{Z98} (also \cite{ZM98}). This discovery opened a way to classification of N-orthogonal surfaces \cite{Z98}.

In the present work we are dealing with dispersionless integrability, which is somewhat technically
different from the standard case (with dispersion). 
However, many dispersionless integrable systems are
also connected with geometry, and reductions, including differential reductions, play an important
role. The integrable system we consider here was introduced to the literature rather recently. First,
the Lax pair corresponding to this system appeared in the work \cite{BDM07} as an unreduced form of
the Lax pair for the Dunajski system, generalising the second heavenly equation \cite{Dun02}. 
And later
\cite{DFK15} it was demonstrated that the system of equations related to this Lax pair describes a general
local form of self-dual conformal structure (SDCS) for the signature (2,2), thus in this work
we call it SDCS system. In integrability  framework  this system possesses a rich structure 
of reductions, including differential reductions. We will introduce several characteristic reductions
for this system, using the Lax pair, hierarchy structure and the dressing scheme. We use reductions to construct solutions for the SDCS system.
We hope that, due to the geometric meaning of the system, these reductions will find 
some geometric interpretation and could be useful for the study of self-dual conformal structures.
We will concentrate on the less known and new reductions, not paying much attention to 
the well known Dunajski
system and second heavenly equation. 

One of our goals is to present type B SDCS system and
consider its reductions and relations with the SDCS system, which appear to be rather
unexpected. The term type B  here refers to BKP hierarchy, where it is connected with a type B
infinite-dimensional Lie algebra. In dispersionless case the symmetry characterising the 
BKP hierarchy is manifested as a symmetry of wave functions with respect to reflection
of the spectral parameter, it is preserved only by odd times of the initial hierarchy, similar 
to BKP hierarchy (see e.g. \cite{BK2005}).
\section{SDCS system of equations}
Let us consider the Lax pair \cite{BDM07}
\bea
\begin{aligned}
X_1&=\p_z-\lambda\p_x
+F_x\p_x+G_x\p_y+ f_x\p_\lambda,
\\
X_2&=\p_w- \lambda\p_y + F_y\p_x+G_y\p_y+f_y\p_\lambda.
\label{LaxSDCS}
\end{aligned}
\eea
This Lax pair represents vector fields of the lowest order (in $\lambda$) 
of SDCS  system hierarchy (see Section \ref{Hi} for more detail).
Commutation relations for the Lax pair give a
coupled system of three second-order nonlinear PDEs for functions $F$, $G$, $f$
\bea
\left\{
\begin{aligned}
Q(F)&=f_y,
\\
Q(G)&=-f_x,
\\
Q(f)&=0,
\end{aligned}
\right.
\label{SDCS3}
\eea
where a linear second order differential operator $Q$ is expressed as
\bea
\begin{aligned}
Q&=(\p_w+F_y\p_x+G_y\p_y)\p_x- (\p_z+F_x\p_x+G_x\p_y)\p_y
\\
&=\p_w\p_x-\p_z\p_y+F_y{\p_x}^2-G_x{\p_y}^2-
(F_x-G_y)\p_x\p_y.
\end{aligned}
\label{Q}
\eea
System (\ref{SDCS3}) can be rewritten in the form of a
coupled system of third order PDEs for the functions $F$, $G$
\bea
\left\{
\begin{aligned}
&
\p_x(Q(F))+\p_y(Q(G))=0,
\\
&
(\p_w+F_y\p_x+G_y\p_y)Q(G)+(\p_z+F_x\p_x+G_x\p_y)Q(F)=0,
\label{sd_3rd}
\end{aligned}
\right.
\label{SDCS2}
\eea
in this form it was introduced in \cite{DFK15}  as a general
local form of equations describing an (anti)self-dual conformal structure (SDCS) for the signature (2,2),
which is locally represented by a metric
\bea
\label{ASDmetric1}
g=dwdx-dzdy-F_y dw^2-(F_x-G_y)dwdz + G_xdz^2,
\eea
while the operator $Q$ defines a symmetric bivector corresponding to the conformal
structure.
\subsection*{Interpolating reductions}
In our consideration of interpolating reductions we follow work \cite{LVB11}.
Let us introduce a formally adjoint Lax pair
\bea
\begin{aligned}
-X_1^*&=\p_z-\lambda\p_x
+F_x\p_x+G_x\p_y+ f_x\p_\lambda + (F_{xx}+G_{xy}),
\\
-X_2^*&=\p_w- \lambda\p_y 
+ F_y\p_x+G_y\p_y+f_y\p_\lambda + (F_{xy}+G_{yy}),
\label{LaxSDCSadj}
\end{aligned}
\eea
which is no longer represented by pure vector fields and contains terms with a multiplication
by the function (a divergence of the respective vector field). However, the compatibility conditions
remain the same and give system (\ref{SDCS3}).
A simplest interpolating reduction is defined by the existence of a wave function of the form 
$\Psi^*=\exp(-\alpha\lambda)$ for the adjoint operators,
\bea
X_{1}^*\exp(-\alpha\lambda)=0, \quad X_{2}^*\exp(-\alpha\lambda)=0.
\label{int}
\eea
For $\ln \Psi^*$ we have nonhomogeneous linear equations
\bea
X_1 \ln \Psi^* + (F_{xx}+G_{xy})=0,\quad X_2 \ln \Psi^* + (F_{xy}+G_{yy})=0.
\label{adjnonhom}
\eea
The reduction condition (\ref{int}) implies the relation
\bea
\alpha f= F_{x}+G_{y},
\label{intdif}
\eea
which represents a differential reduction for SDCS system (\ref{SDCS3}).
Under this condition system (\ref{SDCS3}) reduces to the system of two 
second order equations
\bea
\left\{
\begin{aligned}
Q(F)&=\alpha^{-1}(F_{xy}+G_{yy}) ,
\\
Q(G)&=-\alpha^{-1} (F_{xx}+G_{xy}).
\end{aligned}
\right.
\label{SDCSint}
\eea
The limit of this system for $\alpha\rightarrow 0$ gives the Dunajski system,
defined by the reduction condition $F_{x}+G_{y}=0$, and the limit $\alpha\rightarrow \infty$
leads to linearly degenerate system with $f=0$. Thus, by analogy with the  
interpolating system introduced in \cite{Dun08}, 
we call system (\ref{SDCSint}) the interpolating SDCS system.
\subsubsection*{Higher order interpolating reductions}
The reduction condition for higher interpolating reductions is that
adjoint linear operators (\ref{LaxSDCSadj}) admit a wave function 
of the form $\exp(- \alpha(\lambda^n + P_{n-2}))$,
where $ P_{n-2}$ is a  $\lambda$ polynomial of degree $n-2$, in terms of nonhomogeneous
equations (\ref{adjnonhom}) we have a polynomial solution $- \alpha(\lambda^n + P_{n-2})$. This  condition
defines a reduction for the whole hierarchy (see Section \ref{Hi}).  Second order interpolating reduction is represented by differential relations
\bea
(\p_z
+F_x\p_x+G_x\p_y)f +
\frac{1}{2\alpha}(F_{xx}+G_{xy})=0,
\nn\\
(\p_w + F_y\p_x+G_y\p_y)f
+\frac{1}{2\alpha}(F_{xy}+G_{yy})=0.
\label{int2}
\eea
\subsection*{A simple non-autonomous reduction}
Let Lax pair (\ref{LaxSDCS}) admit a wave function $x+\lambda z$ (see \cite{BDM07}, also \cite{YS15}).
For higher flows of the hierarchy this condition requires modification 
and becomes more sophisticated (see Section \ref{Hi}).
The reduction condition is equivalent to the relation 
\bea
F=-zf.
\label{cond1}
\eea
Under this condition system (\ref{SDCS3}) reduces to the system of two 
second order equations for $f$ and $G$. This reduction becomes rather evident if we rewrite
system (\ref{SDCS3}) in the form
\bea
\left\{
\begin{aligned}
Q(F+zf)&=0,
\\
Q(G + wf)&=0,
\\
Q(f)&=0,
\end{aligned}
\right.
\label{SDCS3'}
\eea
We can simultaneously require a second analogous reduction corresponding to existence 
of the wave function $y+\lambda w$, leading to the condition
\bea
G=-wf.
\label{cond2}
\eea
Under two conditions (\ref{cond1}), (\ref{cond2})  system (\ref{SDCS3}) reduces to one
secong order equation for  $f$ (see \cite{YS15}),
\bea
f_{wx}-f_{zy} - z f_y f_{xx} +w f_x f_{yy}+( zf_x - wf_y) f_{xy}=0
\label{f}
\eea
and conformal structure (\ref{ASDmetric1})  takes the form
\bea
\label{ASDmetricf}
g=dwdx-dzdy
+z f_y dw^2
+(zf_x-w f_y)dwdz 
- w f_xdz^2,
\eea

If we consider the case of divergence-free vector fields (volume-preserving flows),
we also get  an extra condition for $f$,
\bea
zf_x+wf_y=0,
\label{f0}
\eea
which reduces equation (\ref{f}) to the relation
\bea
wf_w + zf_z + f=0.
\label{f1}
\eea
It is easy to check directly that solutions to both relations (\ref{f0}), (\ref{f1})
(and thus to equation (\ref{f}))
are given by the formula
\bea
f(x,z,y,w)=
\frac{1}{2\pi i}\oint R(x +\lambda z;y+\lambda w) d\lambda,
\label{solSD}
\eea 
where $R$ is an arbitrary analytic function of two complex variables
($R(x +\lambda z;y+\lambda w)$ should be $\lambda$ analytic in the neighbourhood
of the integration contour). This formula can be obtained from the dressing scheme,
see Section \ref{Hi}. 
It defines a special self-dual conformal structure by expression (\ref{ASDmetricf}).

Let us consider an example 
\bea
R(x +\lambda z;y+\lambda w) =\sum_i \frac{\psi_i(y+\lambda w)}{x +\lambda z - b_i}
+ \sum_i \frac{\phi_i(x +\lambda z)}{y+\lambda w - c_i}
\label{special}
\eea
where $\psi_i(\lambda)$,   $\phi_i(\lambda)$ are arbitrary functions analytic
in the complex plane, $b_i$, $c_i$ are arbitrary constants, and the contour of integration
in (\ref{solSD}) is a circle of a big radius. Then from formula (\ref{solSD}) we obtain
\bea
f=\sum_i z^{-1}\psi_i(z^{-1}(yz - xw + b_i w)) + \sum_i w^{-1}\phi_i(w^{-1}(xw - yz + c_i z)) 
\eea 
This expression defines a simple explicit example of 
self-dual conformal structure (\ref{ASDmetricf}) depending on arbitrary number of functions of 
one variable. A more involved example is considered below in Section \ref{Hi}.

Instead of divergence-free condition,  we can complement a pair 
of reductions (\ref{cond1}), (\ref{cond2})
with other reduction. Let us consider first  a second order interpolating reduction (\ref{int2}),
which leads  to dimensional reduction for equation (\ref{f}), defining its solutions through 
solutions of two compatible three-dimensional equations
\bea
f_z
-z f_x f_x -wf_x f_y -
\frac{1}{2\alpha}(zf_{xx}+wf_{xy})=0,
\nn\\
f_w - zf_y f_x - w f_y f_y
+\frac{1}{2\alpha}(zf_{xy}+wf_{yy})=0.
\label{int2e}
\eea

Another interesting opportunity to complement a pair of reductions (\ref{cond1}), (\ref{cond2})
and perform a reduction to three-dimensional case is to consider a waterbag ansatz 
for the wave function $\Psi$,
\beaa
\Psi = \lambda + \sum_i c_i \ln(\lambda- u_i).
\eeaa
Substituting this ansatz to the Lax pair (\ref{LaxSDCS}), we get
\beaa
\begin{aligned}
(\p_z +F_x\p_x+G_x\p_y)u_i -u_i\p_x u_i - f_x =0
\\
(\p_w +F_y\p_x+G_y\p_y)u_i -u_i\p_y u_i - f_y =0
\\
f=-\sum_i c_i u_i.
\end{aligned}
\eeaa
Then, taking into account reductions (\ref{cond1}), (\ref{cond2}), we obtain two
closed systems of non-autonomous three-dimensional equations for the functions $u_i$,
\beaa
(\p_z - z f_x\p_x -w f_x\p_y)u_i -u_i\p_x u_i - f_x =0,
\\
(\p_w - z f_y\p_x - w f_y\p_y)u_i -u_i\p_y u_i - f_y =0,
\\
f=-\sum_i c_i u_i.
\eeaa
A common solution of these systems defines a solution of system (\ref{f}) and, respectively,
a self-dual conformal structure (\ref{ASDmetricf}) by the formula $f=-\sum_i c_i u_i$.
\subsection*{Reflection type symmetries and nonlocal SDCS equations}
Reflection type symmetries were introduced to integrable systems theory 
rather recently \cite{AbM13}
and have been extensively studied, especially for a number of systems connected 
with the nonlinear Schrödinger equation.  These symmetries lead to nonlocal systems
with reflection symmetry (P, T, PT type). For dispersionless integrable systems,
as far as the author knows, this type of symmetries was not yet considered. The reason is
probably that mostly systems with deep reductions, having the symplest structure, were
studied. However, the SDCS system possesses rich enough structure of reductios to
admit reflection type symmetry, leading to nonlocal equations.
 
Let us consider a
symmetry condition for SDCS system (\ref{SDCS3})
\bea
G=\wt F,\quad   f=\wt f,     \quad \wt F(x,z;y,w) :=F(y,w;x,z).
\label{reflection}
\eea
We obtain a system of two nonlocal equations
\beaa
\left\{
\begin{aligned}
Q(F)&=f_y,
\\
Q(f)&=0,
\end{aligned}
\right.
\eeaa
where linear second order differential operator $Q$ is given by
\beaa
\begin{aligned}
Q&=\p_w\p_x-\p_z\p_y+F_y{\p_x}^2-\wt F_x{\p_y}^2-
(F_x-\wt F_y)\p_x\p_y.
\end{aligned}
\eeaa
and possesses a symmetry $\wt Q=-Q$.

Respectively, for B-type SDCS system (\ref{SDCS3B}) (see below) we have a system
\beaa
\left\{
\begin{aligned}
Q(F)&=2(\p_w+F_y\p_x+\wt F_y\p_y)f - 4f f_y,
\\
Q(f)&=0,
\end{aligned}
\right.
\eeaa
\section{Type B SDCS system}
Technically this section is close to work  \cite{LVB21},
where we introduced type B Manakov-Santini system.
Type B reduction of SDCS system hierarchy
in terms of vector fields is characterised
by the  symmetry $X(-\lambda)=X(\lambda)$ (see also Section \ref{Hi}).
Lowest order Lax pair reads
\bea
\begin{aligned}
X_1&=\p_{z}-\lambda^2\p_x
+(F_x-2f)\p_x+G_x\p_y+ \lambda f_x\p_\lambda,
\\
X_2&=\p_{w}- \lambda^2\p_y + F_y\p_x+(G_y-2f)\p_y+\lambda f_y\p_\lambda,
\label{BLaxSDCS}
\end{aligned}
\eea
where for simplicity we use the same notations for times and coefficients
as in the Lax pair (\ref{LaxSDCS}), though in the framework of the hierarchy
they correspond to different flows. Commutation relations for the Lax pair
(\ref{BLaxSDCS}) imply the system of equations
\bea
\left\{
\begin{aligned}
Q(F)&=2(\p_w+F_y\p_x+G_y\p_y)f - 4f f_y,
\\
Q(G)&=-2(\p_z+F_x\p_x+G_x\p_y)f  + 4f f_x,
\\
Q(f)&=0,
\end{aligned}
\right.
\label{SDCS3B}
\eea
where operator $Q$ is of the same form as for the SDCS system (\ref{Q}),
\beaa
Q=\p_w\p_x-\p_z\p_y+F_y{\p_x}^2-G_x{\p_y}^2-
(F_x-G_y)\p_x\p_y.
\eeaa
And, rather unexpectedly,{\em functions $F$, $G$ satisfy the same system of third order
equations (\ref{SDCS2})}! It is possible to check it directly, and it will be rather clear 
using the transformation from the B-type Lax pair (\ref{BLaxSDCS}) to the SDCS system 
Lax pair (\ref{LaxSDCS}) that we construct below. Thus, solutions of the 
B-type SDCS system (\ref{SDCS3B}) also give a self-dual conformal structure
by formula (\ref{ASDmetric1}).
\subsection*{Transformation to the SDSC system}
Following the lines of the work \cite{LVB21}, let us perform
a change of the spectral variable in Lax pair (\ref{BLaxSDCS})
$$
\mu=\lambda^2+2f.
$$
As a result, we obtain
\bea
\begin{aligned}
X'_1&=\p_{\wt z}-\mu\p_x
+F_x\p_x+G_x\p_y+ 
2(f_{\wt z} - ff_x + F_xf_x + G_x f_y)\p_\mu,
\\
X'_2&=\p_{\wt w}-\mu\p_y 
+ F_y\p_x+ G_y\p_y+
2(f_{\wt w} - ff_y +F_yf_x + G_y f_y)\p_\mu.
\label{BLaxSDCS'}
\end{aligned}
\eea
Compairing Lax pairs (\ref{LaxSDCS}) and (\ref{BLaxSDCS'}), we get
Miura trnsformation from solutions of B-type SDCS system (\ref{SDCS3B}) to solutions of 
SDCS system (\ref{SDCS3})
\beaa
F\rightarrow F, \quad G\rightarrow G, \quad f=\phi,
\eeaa
where potential $\phi$ is defined through solutions of B-type SDCS system (\ref{SDCS3B})
by the relations
\beaa
\phi_x=2(f_{\wt z} - ff_x + F_xf_x + G_x f_y),
\\
\phi_y=2(f_{\wt w} - ff_y +F_yf_x + G_y f_y).
\eeaa
On the level of the third order system (\ref{SDCS2}) SDCS and type B SDCS systems 
are identical.
\subsection*{Interpolating reduction for type B SDCS system}
To introduce interpolating reductions, we consider a formal conjugation of Lax pair (\ref{BLaxSDCS}),
\beaa
\begin{aligned}
-X_{1}^*&=\p_{z}-\lambda^2\p_x
+(F_x-2f)\p_x+G_x\p_y+ \lambda f_x\p_\lambda +(F_{xx} + G_{xy} -f_x),
\\
-X_{2}^*&=\p_{w}- \lambda^2\p_y + F_y\p_x+(G_y-2f)\p_y+\lambda f_y\p_\lambda
+(F_{xy} + G_{yy} -f_y).
\end{aligned}
\eeaa
A simplest interpolating reduction in this case  is defined 
by existence of the wave function 
$\Psi=\lambda^\alpha$ for adjoint operators (see section \ref{Hi}),
\beaa
X_{1}^*\lambda^\alpha=0, \quad X_{2}^*\lambda^\alpha=0,
\eeaa
that implies
\bea
({1-\alpha})f=( F_{x} + G_{y} ).
\label{intB}
\eea
Under this condition system (\ref{SDCS3B}) reduces to the system of two second order
equations
\bea
\left\{
\begin{aligned}
Q(F)&=\beta(\p_w+F_y\p_x+G_y\p_y)( F_{x} + G_{y} ) 
- \beta^2 ( F_{x} + G_{y} ) ( F_{xy} + G_{yy} ),
\\
Q(G)&=-\beta(\p_z+F_x\p_x+G_x\p_y) ( F_{x} + G_{y} ) 
+ \beta^2  ( F_{x} + G_{y} ) ( F_{xx} + G_{yx} ) ,
\end{aligned}
\right.
\label{SDCS3Bint}
\eea
where $\beta=\frac{2}{1-\alpha}$. 
\subsubsection*{Type B interpolating reduction in terms of the basic SDCS system}
It is possible to characterise  type B interpolating reduction leading to system 
(\ref{SDCS3Bint}) in terms of the basic SDCS system and its Lax pair.
Assuming the existence of a wave function of the form $(\lambda + g)^\gamma$ for the adjoint Lax pair (\ref{LaxSDCSadj}), we obtain the relations
\beaa
&&
\gamma g=F_x + G_y,
\\
&&
f_x=g_{z} - gg_x + F_xg_x + G_x g_y,
\\
&&
f_y=g_{w} - gg_y +F_yg_x + G_y g_y,
\eeaa
which reduce SDCS system  (\ref{SDCS3}) to type B interpolating system
(\ref{SDCS3Bint}), $\gamma=\beta^{-1}$.
\section{The SDCS system hierarchy, the dressing scheme and reductions}
\label{Hi}
To define the hierarchy,
we introduce three series \cite{BDM07}, \cite{LVB11}
\bea
&&
\Psi^0=\lambda+\sum_{n=1}^\infty \Psi^0_n(\mathbf{t}^1,\mathbf{t}^2)\lambda^{-n},
\label{form0}
\\&&
\Psi^1=\sum_{n=0}^\infty t^1_n (\Psi^0)^{n}+
\sum_{n=1}^\infty \Psi^1_n(\mathbf{t}^1,\mathbf{t}^2)\lambda^{-n}
\label{form1}
\\&&
\Psi^2=\sum_{n=0}^\infty t^2_n (\Psi^0)^{n}+
\sum_{n=1}^\infty \Psi^2_n(\mathbf{t}^1,\mathbf{t}^2)\lambda^{-n}.
\label{form2}
\eea 
where $\mathbf{t}^1=(t^1_0,\dots,t^1_n,\dots)$, $\mathbf{t}^2=(t^2_0,\dots,t^2_n,\dots)$.
The SDSC system  hierarchy is defined by the generating relation
\be
(J_0^{-1}\d \Psi^0\wedge \d \Psi^1\wedge \d \Psi^2)_-=0,
\label{analyticity0D}
\ee 
where we use the projectors $(\sum_{-\infty}^{\infty}u_n \lambda^n)_+
=\sum_{n=0}^{\infty}u_n \lambda^n$,
$(\sum_{-\infty}^{\infty}u_n \lambda^n)_-=\sum_{-\infty}^{n=-1}u_n \lambda^n$,
$J_0$ is the determinant of the Jacobian matrix,
\be
J_0=
\begin{vmatrix}
\Psi^0_\lambda &\Psi^1_\lambda&\Psi^2_\lambda \\
\Psi^0_x&\Psi^1_x &\Psi^2_x \\
\Psi^0_y&\Psi^1_y &\Psi^2_y
\end{vmatrix},
\label{J0}
\ee
where $x=t^1_0$, $y=t^2_0$.
Generating relation (\ref{analyticity0D}) implies
Lax-Sato equations of the hierarchy  (see \cite{BDM07}, \cite{LVB11})
\bea
\partial^1_n\mathbf{\Psi}=
+\left(
\frac{(\Psi^0)^{n}}{J_0}
\begin{vmatrix}
\Psi^0_\lambda & \Psi^2_\lambda\\
\Psi^0_y & \Psi^2_y
\end{vmatrix}
\right)_+\partial_x \mathbf{\Psi}-
\left(\frac{(\Psi^0)^{n}}{J_0}
\begin{vmatrix}
\Psi^0_\lambda & \Psi^2_\lambda\\
\Psi^0_x & \Psi^2_x
\end{vmatrix}
\right)_+\partial_y \mathbf{\Psi}-
\nn\\
\left(
\frac{(\Psi^0)^{n}}{J_0}
\begin{vmatrix}
\Psi^0_x & \Psi^2_x\\
\Psi^0_y & \Psi^2_y
\end{vmatrix}
\right)_+\partial_\lambda \mathbf{\Psi}
,
\label{Dun11}
\eea
\bea
\partial^2_n\mathbf{\Psi}=
-\left(\frac{(\Psi^0)^{n}}{J_0}
\begin{vmatrix}
\Psi^0_\lambda & \Psi^1_\lambda\\
\Psi^0_y & \Psi^1_y
\end{vmatrix}
\right)_+\partial_x \mathbf{\Psi}+
\left(\frac{(\Psi^0)^{n}}{J_0}
\begin{vmatrix}
\Psi^0_\lambda & \Psi^1_\lambda\\
\Psi^0_x & \Psi^1_x
\end{vmatrix}
\right)_+\partial_y \mathbf{\Psi}+
\nn\\
\left(\frac{(\Psi^0)^{n}}{J_0}
\begin{vmatrix}
\Psi^0_x & \Psi^1_x\\
\Psi^0_y & \Psi^1_y
\end{vmatrix}
\right)_+\partial_\lambda \mathbf{\Psi},
\label{Dun21}
\eea
where $\mathbf{\Psi}=(\Psi^0, \Psi^1, \Psi^2)$, $\partial^k_n=\frac{\partial}{\partial t^k_n}$.
The first two flows of the hierarchy (\ref{Dun11}), (\ref{Dun21})
read
\bea
&&\p_1^1\mathbf{\Psi}=(\lambda\p_x
-F_{x}\p_x-G_{x}\p_y-f_{x}\p_{\lambda})\mathbf{\Psi},
\label{fl01}
\\
&&\p_1^2\mathbf{\Psi}=(\lambda\p_y
-F_{y}\p_x
-G_{y}\p_y -f_{y}\p_{\lambda})\mathbf{\Psi},
\label{fl02}
\eea
here
$$
G=\Psi^2_1,\quad F=\Psi^1_1, \quad f=\Psi^0_1,
$$
and after the identification $z=t^1_1$, $w=t^2_1$ we obtain linear equations 
for Lax pair (\ref{LaxSDCS}) with  $\Psi^0, \Psi^1, \Psi^2$  
playing a role of the basic wave functions.

The second flows can be represented in a form
\bea
&&\p_2^1\mathbf{\Psi}=
(\lambda \p_1^1+f \p_x
-(\p_1^1 F)\p_x-(\p_1^1 G)\p_y-(\p_1^1f)\p_{\lambda}
)
\mathbf{\Psi},
\label{fl110}
\\
&&\p_2^2\mathbf{\Psi}=
(\lambda \p_1^2+f \p_y
-(\p_1^2 F)\p_x-(\p_1^2 G)\p_y-(\p_1^2f)\p_{\lambda}
)
\mathbf{\Psi}.
\label{fl120}
\eea
To write down vector fields more explicitly,
one should use first flows (\ref{fl01}), (\ref{fl02}). 
\subsection*{The dressing scheme}
Let us consider nonlinear vector  Riemann problem of the form (see \cite{BDM07})
\bea
\mathbf{\Psi}_\text{out}= \mathbf{R}(\mathbf{\Psi}_\text{in}),
\label{Riemann}
\eea
or, more explicitely,
\beaa
&&
\Psi^0_\text{out}=R_0(\Psi^0_\text{in},\Psi^1_\text{in},\Psi^2_\text{in}),
\nn\\
&&
\Psi^1_\text{out}=R_1(\Psi^0_\text{in},\Psi^1_\text{in},\Psi^2_\text{in}),
\nn\\
&&
\Psi^2_\text{out}=R_2(\Psi^0_\text{in},\Psi^1_\text{in},\Psi^2_\text{in}),
\eeaa
where $\mathbf{\Psi}_\text{out}$, $\mathbf{\Psi}_\text{in}$ denote the boundary values
inside and outside the unit circle (or other closed curve) in the complex plane.
The problem is to find a vector function analytic
outside the curve with some fixed singularity at infinity 
which satisfies relation
(\ref{Riemann}). The set of functions $R_0$, $R_1$, $R_2$ defines a complex
diffeomorphism connecting
the boundary values of the vector function $\mathbf{\Psi}$
outside and inside the unit circle, and we call this set a dressing data.

To get solutions of the general hierarchy using the problem (\ref{Riemann}),
one should find solution to this problem with the singulariries at $\lambda=\infty$
defined by the series (\ref{form0}), (\ref{form1}), (\ref{form2})
(for simplicity we suggest that only a final number of times is not equal to zero).
Relation (\ref{Riemann}) implies that the form
$$
\Omega=J_0^{-1}\d \Psi^0\wedge \d \Psi^1\wedge \d \Psi^2
$$
is analytic in the whole complex plane,
and so its expansion at infinity satisfies the generating identity  (\ref{analyticity0D}).
Then the solution of the Riemann problemm (\ref{Riemann}) satisfies
Lax-Sato equations (\ref{Dun11}), (\ref{Dun21}) and provides a solution for the
SDCS system hierarchy. 
\subsection*{Reductions in terms of the hierarchy}
\subsubsection*{Interpolating reduction}
In terms of the SDSC system hierarchy interpolating reductions 
are defined by the relation \cite{LVB11}
\bea
J_0=\exp(\alpha(\Psi^0)^k_-),
\label{intJ}
\eea
which is  equivalent to the condition that $\Psi^*=\exp(-\alpha(\Psi^0)^k_+)$  is a  wave function
for adjoint linear operators of the hierarchy (\ref{Dun11}), (\ref{Dun21}).  We used the existence
of wave functions  of this form to define interpolating reductions (\ref{intdif}), (\ref{int2})
for the SDCS system.

In terms of the dressing scheme interpolating reduction is given by a `twisted' volume-preservation
condition \cite{LVB11} for the diffeomorphism $\mathbf{R}$ (\ref{Riemann}),
\be
\mathbf{f}\circ\mathbf{R}\circ\mathbf{f}^{-1}\in \text{SDiff(N+1)},
\label{redMSdressing}
\ee
where diffeomorphism $\mathbf{f}$ is defined as
\bea
&&
f_0(\mathbf{\Psi})=\Psi^0,
\nn\\&&
f_1(\mathbf{\Psi})=
\exp\left(-\frac{\alpha}{2}(\Psi^0)^k\right)\Psi^1,
\nn\\&&
f_2(\mathbf{\Psi})=
\exp\left(-\frac{\alpha}{2}(\Psi^0)^k\right)\Psi^2 ,
\label{reddiffMS}
\eea

For the type B  SDCS system hierarchy $J_0$ 
possesses a symmetry $J_0(-\lambda)=J_0(\lambda)$,
which is compatible with condition (\ref{intJ}) only for even $k$. However,
in this case it is also possible to define a more elementary interpolating reduction
(which degenerates for the SDCS system) by the relation
\bea
J_0=\left(\frac{\Psi^0}{\lambda}\right)^{-\alpha},
\eea
which implies that $\Psi^*=\lambda^\alpha$ is a wave function of adjoint linear operators
of the hierarchy, and for B type SDSC system we get a reduction condition  (\ref{intB}).
\subsubsection*{Non-autonomous reductions and solutions}
Non-autonomous reductions, related to conditions (\ref{cond1}), (\ref{cond2}),
in terms of the hierarchy are defined as (see \cite{BDM07}, also \cite{YS15})
\bea
(\Psi^1)_-=0, \quad (\Psi^2)_-=0,
\label{condH}
\eea
then $\Psi^1=\left(\sum_{n} t^1_n (\Psi^0)^{n}\right)_+$,
$\Psi^2=\left(\sum_{n} t^2_n (\Psi^0)^{n}\right)_+$, and series
(\ref{form1}), (\ref{form2}) are of the form
\bea
\Psi^1=\sum_{n=0}^{\infty} t^1_n (\Psi^0)^{n} 
- \left(\sum_{n=0}^{\infty} t^1_n (\Psi^0)^{n}\right)_-,
\label{1r}
\\
\Psi^2=\sum_{n=0}^{\infty} t^2_n (\Psi^0)^{n} 
- \left(\sum_{n=0}^{\infty} t^2_n (\Psi^0)^{n}\right)_-.
\label{2r}
\eea
Thus, due to reduction conditions (\ref{condH}), wave functions $\Psi^1$,
$\Psi^2$ are expressed through the wave function $\Psi^0$.  
For higher times equal to zero $\Psi^1=x+\lambda z$,  $\Psi^2=y+\lambda w$,
implying conditions  (\ref{cond1}), (\ref{cond2}).
However, switching on higher times leads to more sophisticated reduction conditions.
Let us suggest that $t^1_2$ not equal to zero and consider first of conditions  (\ref{condH}),
which implies
\beaa
\Psi^1=t^1_2(\lambda^2 + 2f) +z\lambda + x.
\eeaa
Substituting this expression to Lax pair (\ref{LaxSDCS}), we get differential reduction
\beaa
F_x + z f_x +2 t^1_2(f_z+F_x f_x+ G_x f_y)=0,
\\
F_y + z f_y +2 t^1_2(f_w+ F_y f_x+ G_y f_y)=0.
\eeaa
In terms of Riemann problem (\ref{Riemann}) reductions (\ref{condH}) mean that the functions
$\Psi^1$, $\Psi_2$ have no discontinuity on the contour, and Riemann problem (\ref{Riemann})
takes the form
\beaa
&&
\Psi^0_\text{out}=R_0(\Psi^0_\text{in},\Psi^1_\text{in},\Psi^2_\text{in}),
\nn\\
&&
\Psi^1_\text{out}=\Psi^1_\text{in},
\nn\\
&&
\Psi^2_\text{out}=\Psi^2_\text{in}.
\eeaa
For a case of volume-preserving diffeomorphism
\bea
\Psi^0_\text{out}=\Psi^0_\text{in} + R_0(\Psi^1,\Psi^2),
\label{redRiemann}
\eea
where, due to the reduction, $\Psi^1$, $\Psi^2$ are $\lambda$ polynomials (for a finite 
number of higher times not equal to zero), with the coefficients defined through the coefficients
of the series for $\Psi^0$. Constructing a solution to the problem (\ref{redRiemann}), 
\bea
\Psi^0 (\lambda)=
\frac{1}{2\pi i}\oint \frac{R_0(\Psi^1(\mu),\Psi^2(\mu))}{\lambda-\mu} d\mu,
\label{solRiem}
\eea
it is possible
to find the coefficients of expansion of  $\Psi^0$ in terms of implicit functions 
and construct a solution
to SDCS system (\ref{SDCS3}). In the simplest case, when higher times are equal to zero,
we have $\Psi^1=x+\lambda z$,  $\Psi^2=y+\lambda w$,
leading to formula (\ref{solSD}),
\beaa
f(x,z,y,w)=
\frac{1}{2\pi i}\oint R_0(x +\lambda z;y+\lambda w) d\lambda.
\eeaa 
A special choice of the function $R_0$ (\ref{special}) allows us to get rid of the
integration and get some solutions to the SDCS system in explicit form.

For nonzero time $t^1_2$, expressions for wave functions take the form
 \beaa
 &&
 \Psi^1=t^1_2(\lambda^2 + 2f) +z\lambda + x,  
 \\
 &&
 \Psi^2=y+\lambda w,
 \eeaa
and formula (\ref{solRiem}) reads
\bea
\Psi^0 (\lambda)=
\frac{1}{2\pi i}\oint 
\frac{R_0(t^1_2(\mu^2 + 2f) +z\mu + x;\;y+\mu w)}{\lambda-\mu} d\mu.
\label{solRiem1}
\eea
By this formula the function $\Psi^0 (\lambda)$ is defined through $f$, which is its first coefficient 
of expansion at infinity, and $f$ is defined in terms of implicit function,
\bea
f(x,z,t^1_2;y,w)=
\frac{1}{2\pi i}\oint 
{R_0(t^1_2(\mu^2 + 2f(x,z,t^1_2;y,w)) +z\mu + x;\;y+\mu w)} d\mu.
\label{implicit}
\eea
Function $G$ for the SDCS system is expressed as $G=-wf$ (\ref{cond2}). Expression for 
function $F$
is given by the $\lambda^{-1}$ coefficient of the series (\ref{form1}). According to 
(\ref{1r}),
\bea
F=-zf - t^1_2 \Psi^0_2,
\label{Fimpl}
\eea
where $\Psi^0_2$ is the $\lambda^{-2}$ coefficient of the series (\ref{form0}),
which can be obtained from formula  (\ref{solRiem1}),
\bea
\Psi^0_2=\frac{1}{2\pi i}\oint 
{R_0(t^1_2(\mu^2 + 2f(x,z,t^1_2;y,w)) +z\mu + x;\;y+\mu w)}\mu d\mu.
\label{Fimpl1}
\eea
Thus we have constructed a solution $f, F, G$ to SDCS system (\ref{SDCS3}), where
$f$ is defined as implicit function (\ref{implicit}), $G=-wf$ and $F$ is defined through $f$
by (\ref{Fimpl}), (\ref{Fimpl1}). This solution depends on an arbitrary analytic function
of two variables $R_0$ and a parameter $t^1_2$.

Let us consider
\beaa
R_0(\Psi^1, \Psi^2)=\sum_i \frac{\phi_i(\Psi^1)}{\Psi^2-c_i} 
\eeaa
where $\phi_i(\lambda)$ are arbitrary functions analytic
in the complex plane, $c_i$ are arbitrary constants. Then it is possible to perform
integration in formulae (\ref{implicit}), (\ref{Fimpl1}) and obtain
\beaa
&&
f(x,z,t^1_2;y,w)=\sum_i w^{-1}
\phi_i \left(t^1_2((\frac{c_i-y}{w})^2 + 2f) +z\frac{c_i-y}{w} + x\right),
\\
&&
\Psi^0_2=\sum_i \frac{c_i-y}{w^2}
\phi_i \left(t^1_2((\frac{c_i-y}{w})^2 + 2f) +z\frac{c_i-y}{w} + x\right).
\eeaa
The solution to SDCS system in this case depends on a set of analytic functions
of one variable  $\phi_i$, constants $c_i$ and a parameter $t^1_2$ (higher time
of the hierarchy).
\subsubsection*{Type B hierarchy}
Type B SDCS system hierarchy is defined by the symmetry condition 
for series (\ref{form0}), (\ref{form1}), (\ref{form2}),
\beaa
\Psi^0(-\lambda)=-\Psi^0(\lambda),\quad \Psi^1(-\lambda)=\Psi^1(\lambda),\quad
\Psi^2(-\lambda)=\Psi^2(\lambda),
\eeaa
which is compatible only with odd flows of the initial hierarchy, 
and vector fields of the hierarchy possess the symmetry  $X(-\lambda)=X(\lambda)$.
Lax pair (\ref{BLaxSDCS})
corresponds to the symmetry reductions of the second flows (\ref{fl110}), (\ref{fl120})
with the identification
$z=t^1_2$, $w=t^2_2$.

In terms of the dressing scheme reduction of type B requires a symmetry
\beaa
&&
R_0(-\Psi^0,\Psi^1,\Psi^2)=-R_0(\Psi^0,\Psi^1,\Psi^2),
\\
&&
R_1(-\Psi^0,\Psi^1,\Psi^2)=R_1(\Psi^0,\Psi^1,\Psi^2),
\\
&&
R_2(-\Psi^0,\Psi^1,\Psi^2)=R_2(\Psi^0,\Psi^1,\Psi^2).
\eeaa
\subsubsection*{Reflection type symmetries}
In terms of the hierarchy reflection type symmetry (\ref{reflection})
is defined by the following conditions for series (\ref{form0}), (\ref{form1}), (\ref{form2}),
\beaa
\Psi^0(\mathbf{t}^1,\mathbf{t}^2)=\Psi^0(\mathbf{t}^2,\mathbf{t}^1),
\quad
\Psi^2(\mathbf{t}^1,\mathbf{t}^2)=\Psi^1(\mathbf{t}^2,\mathbf{t}^1).
\eeaa
The dressing data in this case require a symmetry
\beaa
&&
R_0(\Psi^0,\Psi^1,\Psi^2)=R_0(\Psi^0,\Psi^2,\Psi^1),
\\
&&
R_2(\Psi^0,\Psi^1,\Psi^2)=R_1(\Psi^0,\Psi^2,\Psi^1).
\eeaa


\begin{thebibliography}{99}

\bibitem{Z98}
V.E. Zakharov, Description of the n-orthogonal curvilinear coordinate systems and Hamiltonian integrable systems of hydrodynamic type, I: Integration of the Lame equations, Duke Math. J., 94(1), 103-139, (1998).

\bibitem{ZS74}
V.E. Zakharov, A.B. Shabat, A scheme for integrating the nonlinear equations of mathematical physics by the method of the inverse scattering problem. I, Funk. Anal. Prilozh., 8(3), 43-53 (1974) [Funct. Anal. Appl., 8(3), 226-235, (1974)].

\bibitem{ZM98}
V.E. Zakharov and S.V.Manakov, Reductions in systems integrable by the method of inverse scattering problem, Doklady Matematics, 57(3), 471-474, (1998).

\bibitem{BDM07}
L. V. Bogdanov, V. S. Dryuma and S. V. Manakov,
Dunajski generalization of the second heavenly equation:
dressing method and the hierarchy,
J Phys. A: Math. Theor. \textbf{40} (2007) 14383--14393

\bibitem{Dun02}  M. Dunajski, 
Anti-self-dual four–manifolds with a parallel real spinor,
{Proc. Roy. Soc. Lond. A}
\textbf{458}, 1205 (2002)


\bibitem{DFK15}M. Dunajski, E.V. Ferapontov and 
B. Kruglikov, On the Einstein-Weyl 
and conformal self-duality equations,
Journal of Mathematical Physics 56(8), 083501 (2015).

\bibitem{BK2005}L.V. Bogdanov and B.G. Konopelchenko, On dispersionless BKP hierarchy and its reductions, J. Nonlinear Math. Phys., 12, Suppl.1, 64-73 (2005)

\bibitem{LVB11}
L.V. Bogdanov, Interpolating differential reductions of 
multidimensional integrable hierarchies, 
Theor. Math. Phys., 167(3), 705-713 (2011)

\bibitem{Dun08}
Maciej Dunajski, 
An interpolating dispersionless integrable system, 
J. Phys. A: Math. Theor. 41 (2008) 315202

\bibitem{YS15}
G. Yi and P.M. Santini, 
The inverse spectral transform for the Dunajski hierarchy and some of its reductions: I. Cauchy problem and longtime behavior of solutions,
J. Phys. A: Math. Theor. 48 215203 (2015)

\bibitem{AbM13}
M. J. Ablowitz and Z. H. Musslimani,
Integrable nonlocal nonlinear Schrödinger equation, 
Physical review letters, 110(6), 064105  (2013). 

\bibitem{LVB21}
L.V. Bogdanov, 
Dispersionless BKP equation, the Manakov-Santini system and Einstein-Weyl structures, Symmetry, 13(9), 1699 (2021)

\end{thebibliography}
\end{document}